\documentclass{aa}

\usepackage{graphics}

\begin{document}

\thesaurus{08          
           (08.01.03;  
            08.06.03;  
            08.12.02;  
            08.09.02)  
}

\title{On the Interpretation of the Optical Spectra of L-type Dwarfs\thanks{
          Based on observations made with the William Herschel 
          Telescope (WHT) operated on the island of La Palma 
          by the Isaac Newton Group at the Observatorio del Roque de los 
          Muchachos of the Instituto de Astrof\'\i sica de Canarias; and 
          on observations obtained at the W. M. Keck Observatory, which is 
          operated as a scientific partnership among the California 
          Institute of Technology, the University of California and the 
          National Aeronautics and Space Administration. This observatory 
          was made possible by the generous financial support of the W. M. 
          Keck Foundation.}}

\author{Ya. Pavlenko\inst{1}, M.\,R. Zapatero Osorio\inst{2}, 
       \and
       R. Rebolo\inst{2,3}
       }

\offprints{R. Rebolo, M. Zapatero Osorio, Ya. Pavlenko}
\mail {rrl@iac.es,mosorio@iac.es,yp@mao.kiev.ua}

\institute{Main Astronomical Observatory, Golosiiv, 03680 Kyiv-127, Ukraine
      \and Instituto de Astrof\'\i sica de Canarias. C/. V\'\i a L\'actea 
           s/n, E-38200 La Laguna, Tenerife. Spain
      \and Consejo Superior de Investigaciones Cient\'\i ficas, Spain}

\date{Received ; accepted }

\authorrunning{Pavlenko, Zapatero Osorio, Rebolo}

\titlerunning{On the interpretation of the optical spectra of L-dwarfs}

\maketitle

\begin{abstract}
We present synthetic optical spectra in the red and far-red (640--930\,nm) 
of a sample of field L dwarfs suitably selected to cover this new spectral 
class, and the brown dwarf GL\,229B. We have used the recent ``dusty'' 
atmospheres by Tsuji (\cite{tsuji00}) and by Allard (\cite{allard99}), 
and a synthesis code (Pavlenko et al. \cite{pav95}) working under LTE 
conditions which considers the chemical equilibrium of more than 100 
molecular species and the detailed opacities for the most relevant bands. 
Our computations show that the alkali elements Li, Na, K, Rb, and Cs govern 
the optical spectra of the objects in our sample, with Na and K 
contributing significantly to block the optical emergent radiation. 
Molecular absorption bands of oxides (TiO and VO) and hydrides (CrH, 
FeH and CaH) also dominate at these wavelengths in the early L-types 
showing a strength that progressively decreases for later types. We 
find that the densities of these molecules in the atmospheres of our 
objects are considerably smaller by larger factors than those predicted 
by chemical equilibrium considerations. This is consistent with Ti and 
V atoms being depleted into grains of dust. 

In order to reproduce the overall shape of the optical spectra of our 
observations an additional opacity is required to be implemented in the 
computations. We have modelled it with a simple law of the form 
$a_{\circ} \ (\nu / \nu_{\circ})^N$, with $N$\,=\,4, and found that this 
provides a sufficiently good fit to the data. This additional opacity 
could be due to molecular/dust absorption or to dust scattering. We 
remark that the equivalent widths and intensities of the alkali lines 
are highly affected by this opacity. In particular, the lithium resonance 
line at 670.8\,nm, which is widely used as a substellarity discriminator, 
is more affected by the additional opacity than by the natural depletion 
of neutral lithium atoms into molecular species. Our theoretical spectra 
displays a rather strong resonance feature even at very cool effective 
temperatures ($\sim$1000\,K); depending on the effective temperature 
and on the amount of dust in the atmospheres of very cool dwarfs, it 
might be possible to achieve the detection of lithium even at 
temperatures this cool. Changes in the 
physical conditions governing dust formation in L-type objects will cause 
variability of the alkali lines, particularly of the shorter wavelength 
lines. 
\end{abstract}



\begin{table*}
\caption[]{\label{tab1} Log of spectroscopic observations}
\begin{center}
\small
\begin{tabular}{llccccc}
\hline
Object               & Spec. Type & Telescope & Dispersion & $\Delta \lambda$ & Date   & Exposure \\
                     &            &           & (\AA/pix)  & (nm)             & (1997) & (s)\\
\hline
Kelu\,1              & L2 (L2)    & WHT       & 2.9        & 640--930         & Jun 16 & 2$\times$1200 \\
Denis-P\,J1228--1547 & L4.5 (L5)  & WHT       & 2.9        & 640--930         & Jun 14 & 2$\times$1200 \\
Denis-P\,J0205--1159 & L5 (L7)    & WHT       & 2.9        & 640--930         & Aug 23 & 2$\times$1800 \\
                     &            & KeckII    & 2.5        & 640--890         & Nov 2  & 1$\times$1600 \\
\hline
 & & & & & & \\
\end{tabular}
\end{center}
NOTES. Spectral types are given in Mart\'\i n et al. (\cite{martin99}). 
Those spectral types in brackets come from Kirkpatrick et al. 
(\cite{kirk99a}).
\end{table*}

\section{Introduction}

The optical spectra of the recently discovered very cool dwarfs present 
new challenges to theoretical interpretation. Their spectral 
characteristics are drastically different from those of the well known 
M-dwarfs and this has prompted the use of a new spectral classification, 
the so-called L-dwarfs (Mart\'\i n et al. \cite{martin97a}, \cite{martin99}; Kirkpatrick et al. \cite{kirk99a}). In principle, the study of the optical spectral energy distribution may allow a better understanding of their physical properties, effective temperatures, gravities, atmospheric composition, etc. The main molecular absorbers at optical wavelengths in early- to mid-M dwarfs are significantly depleted at the lower temperatures present in L dwarfs because of the incorporation of their atoms into dust grains. This process should start in the latest M-dwarf atmospheres (Lunine et al. \cite{lunine89}; Tsuji, Ohnaka \& Aoki \cite{tsuji96a}; Tsuji et al. \cite{tsuji96b}; Jones \& Tsuji \cite{jones97}; Allard et al. \cite{allard97}), considerably reducing the strength of TiO, VO and other molecular bands, and producing significant changes in the overall properties of the optical spectra. Naturally, the appearance of dust will modify the temperature structure of the atmosphere, significantly affecting the formation of the emerging spectrum (Allard et al. \cite{allard97}). 

In this paper we follow a semi-empirical approach to understand the relevance of different processes on the resulting spectral energy distributions in the optical for L dwarfs and Gl\,229B. We have obtained far-red optical spectra of several of these cool dwarfs and compared them to synthetic spectra generated 
using the latest models by Tsuji (\cite{tsuji00}) and Allard (\cite{allard99}). These models have been successfully used in the interpretation of the near-infrared spectra of these objects (Tsuji, Ohnaka \& Aoki \cite{tsuji99}; Kirkpatrick et al. \cite{kirk99b}).We have taken into account the depletion of some relevant molecules associated with the formation of dust  and we have investigated the effects of dust scattering and/or absorption on the formation of the optical spectra. Remarkably strong alkali lines are present in the spectra and dominate its shape in the 600--900\,nm region, providing major constraints to the theoretical modelling. We present the observations in section 2, models and synthetic spectra in section 3, and the role of alkalis, molecular bands and dust scattering and/or absorption is considered in section 4. In section 5 we discuss the implications of our study on the determination of effective temperatures and gravities as well as the formation of the lithium resonance line which is a key discriminator of substellar nature in these objects. Finally, conclusions are presented in Section 6.

\section{Observations and data reduction}
  
\begin{figure*}
\begin{center}
\resizebox{16.5cm}{!}{\includegraphics{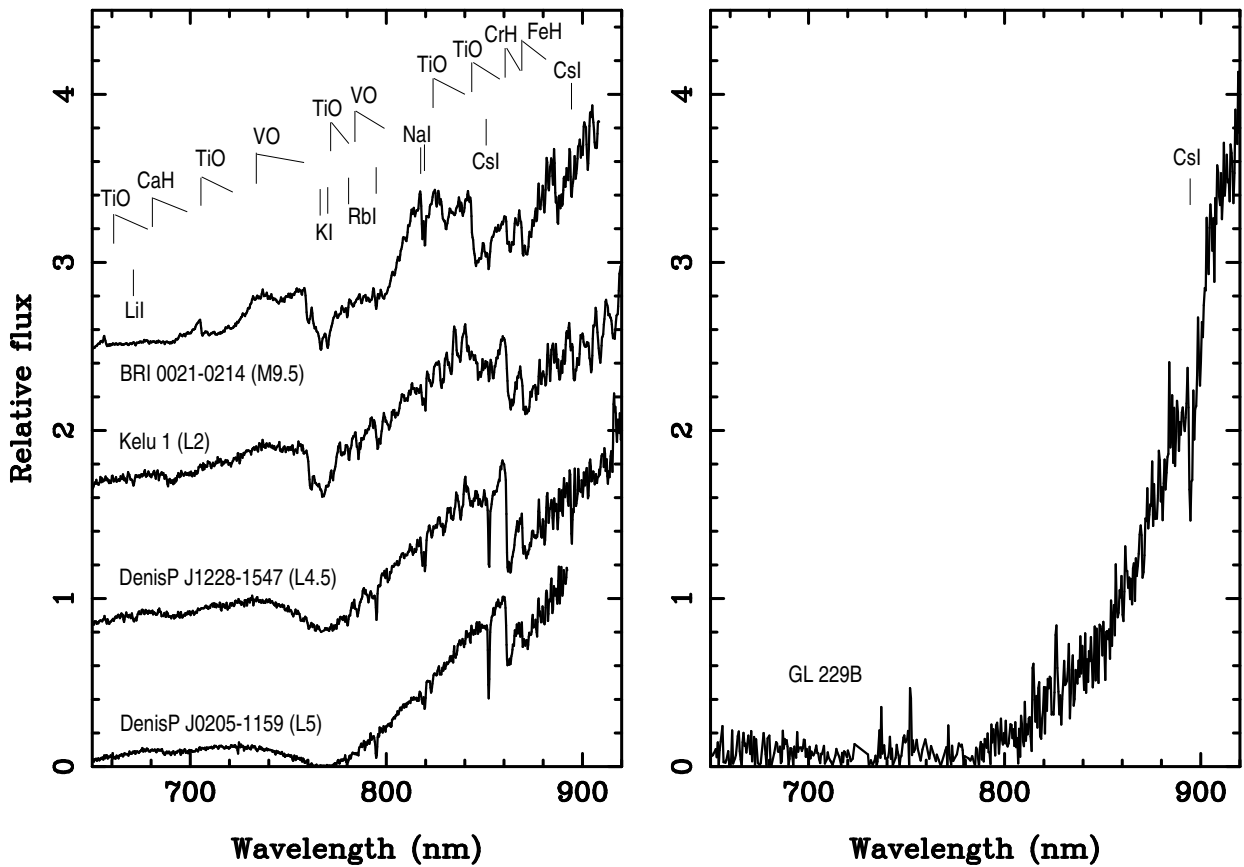}}
\end{center}
\caption[]{\label{fig1} Far-red optical spectra for our sample. 
New observations are 
those of Kelu\,1, Denis-P\,J1228--1547 and Denis-P\,J0205--1159 (KeckII 
spectrum). Data for BRI\,0021-0214 and Gl\,229B have been collected from 
Mart\'\i n, Rebolo \& Zapatero Osorio (1996) and from Schultz et al. 
(1998), respectively. Spectral types for the L-dwarfs are given following 
the classification of Mart\'\i n et al. (\cite{martin99}). Spectra in 
the left panel have been shifted by 0.8 units for clarity. Identification 
of some atomic and molecular features is provided in the top.}
\end{figure*}

We have collected intermediate-resolution spectra in the 640--930\,nm range for Kelu\,1 (Ruiz, Leggett \& Allard \cite{ruiz97}), Denis-P\,J1228--1547 and Denis-P\,J0205--1159 (Delfosse et al. \cite{delfosse97}) using the 4.2\,m William Herschel Telescope (WHT; Observatorio del Roque de los Muchachos, La Palma). The coolest L dwarf in our sample, Denis-P\,J0205--1159, has also been observed with the KeckII telescope (Mauna Kea Observatory, Hawaii). Table~\ref{tab1} summarizes the log of the observations. The instrumentation used was the ISIS double-arm spectrograph at the WHT (only the red arm) with the grating R158R and a TEK (1024$\times$1024\,pix$^2$) CCD detector, and the Low Resolution Imaging Spectrograph (LRIS, Oke et al. \cite{oke95}) with the 600 grove\,mm$^{-1}$ grating and the TEK (1024$\times$1024\,pix$^2$) CCD detector at the KeckII telescope. The nominal dispersion and the wavelength coverage provided by each instrumental setup were similar and are listed in Table~\ref{tab1}. Slit projections were typically 2--3\,pix giving spectral resolutions of 6--8\,\AA. Spectra were reduced by a standard procedure using IRAF\footnote{IRAF is distributed by National Optical Astronomy Observatories, whcih is operated by the Association of Universities for Research in Astronomy, Inc., under contract with the National Science Foundation.}, which included debiasing, flat-fielding, optimal extraction, and wavelength calibration using the sky lines appearing in each individual spectrum (Osterbrock et al. \cite{osterbrock96}). Finally, the observed WHT spectra were corrected from instrumental response making use of the spectrophotometric standard stars BD\,+26$^\circ$\,2606 and HD\,19445 which have absolute flux data available in the IRAF environment. No flux standards were observed at the KeckII telescope, and thus the instrumental signature of the KeckII spectrum of Denis-P\,J0205--1159 was removed by matching it with the WHT spectrum. Our observations are displayed in Fig.~\ref{fig1} together with spectra of the other two objects in our sample: BRI\,0021--0214 (M9.5, Irwin, McMahon, \& Reid \cite{irwin91}), and Gl\,229B (a brown dwarf with methane absorptions, Nakajima et al. \cite{nakajima95}; Oppenheimer et al. \cite{oppenheimer95}). Our sample covers a wide range of spectral types, from the transition objects between M and L types down to the latest types. 

\section{Model atmospheres, spectral synthesis code, chemical equilibrium 
         and opacities}

\begin{figure}
\begin{center}
\resizebox{7.5cm}{!}{\includegraphics{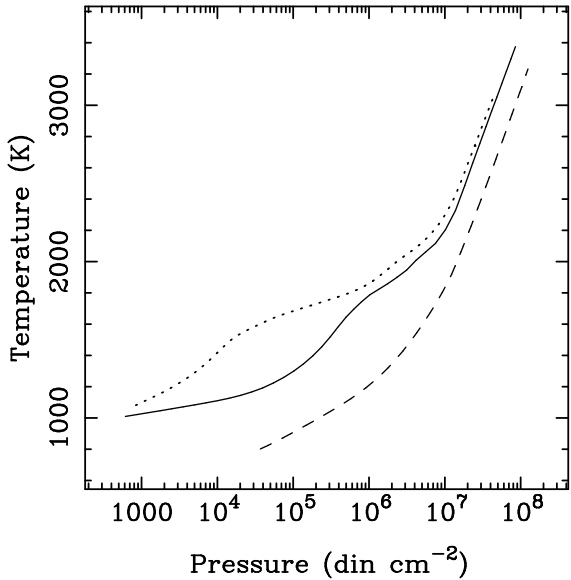}}
\end{center}
\caption[]{\label{fig2} Temperature structures for the model atmospheres 
of $T_{\rm eff}$\,=\,1400\,K and log\,$g$\,=\,5.0 given by Tsuji 
(\cite{tsuji00}, C-type, dashed line; B-type, dotted line) and by Allard 
(\cite{allard99}, full line).}
\end{figure}

We carried out the computations using the LTE spectral synthesis program 
WITA5, which is a modified version of the program used by Pavlenko et al. 
(\cite{pav95}) for the study of the formation of lithium lines in cool 
dwarfs. The modifications were aimed to incorporate ``dusty effects''
which can affect the chemical equilibrium and radiative transfer processes 
in very cool atmospheres. We have used the set of Tsuji's (\cite{tsuji00}) 
``dusty'' (C-type) LTE model atmospheres. These models were computed for 
the case of segregation of dust--gas phases, i.e. for conditions 
$r_{dust} > r_{crit}$, where $r_{dust}$ is the size of dust particles and 
$r_{crit}$ is critical size corresponding to the gas--dust detailed 
equilibrium state. In a previous study (Pavlenko, Zapatero Osorio \& 
Rebolo \cite{pav+oso+reb99}) we had used Tsuji's (\cite{tsuji00}) B-type 
models which are computed for the case of $r_{dust} = r_{crit}$. In this 
paper we have also used a grid of the NextGen ``dusty'' model atmospheres 
computed recently by Allard (\cite{allard99}). The temperature-pressure 
stratification of Allard's models lie between those of the C-type and 
B-type models of Tsuji (\cite{tsuji00}) as it can be seen in Fig.~\ref{fig2}.

Chemical equilibrium was computed for the mix of $\approx$100 molecular 
species. Alkali-contained species formation processes were considered in 
detail because of the important role of the neutral alkali atoms in the 
formation of the spectra. Constants for chemical equilibrium computations 
were taken mainly from Tsuji (\cite{tsuji73}).

\begin{table*}
\begin{center}
\caption[]{\label{tab2} List of the opacity sources used in our computations}
\begin{tabular}{lll}
\hline\hline
\noalign{\smallskip}
Opacity source & Subprogram & Author \\
\hline
\noalign{\smallskip}
Bound-free absorption of H                       &  HOP     &  
Kurucz (\cite{kurucz93}) \\
H$_{2}^{+}$ absorption                           &  H2PLOP  &  
Kurucz (\cite{kurucz93})  \\
H$^-$ ion absorption                             &  HMINOP  &  
Kurucz (\cite{kurucz93}) \\
Rayleigh scattering by H atoms                   &  HRAYOP  &  
Kurucz (\cite{kurucz93})  \\
He\,{\sc i} bound-free absorption                &  HE1OP   &  
Kurucz (\cite{kurucz93})  \\
He\,{\sc ii} bound-free absorption               &  HE2OP   &  
Kurucz (\cite{kurucz93}) \\
He$^-$ absorption                                &  HEMIOP  &  
Kurucz (\cite{kurucz93}) \\
Rayleigh scattering by He                        &  HERAOP  &  
Kurucz (\cite{kurucz93})  \\
Bound-free absorption ofMg\,{\sc i}, Al\,{\sc i}, Si\,{\sc i}, 
Fe\,{\sc i}, OH and CH         &  COOLOP  &  Kurucz (\cite{kurucz93}) \\
Bound-free absorption of N\,{\sc i}, O\,{\sc i}, Mg\,{\sc ii}, 
Si\,{\sc ii}, Ca\,{\sc i}   &  LUKEOP  &  Kurucz (\cite{kurucz93}) \\
Bound-free absorption of C\,{\sc ii} + N\,{\sc ii} + O\,{\sc ii}  
&  HOTOP   &  Kurucz (\cite{kurucz93}) \\
Electron Thompson scattering by e$^-$            &  ELECOP  &  
Kurucz (\cite{kurucz93}) \\
Rayleigh scattering  by H$_2$                    &  H2RAOP  &  
Kurucz (\cite{kurucz93}) \\
Absorption due to Hydrogen lines,                &  HLINOP  &  
Kurucz (\cite{kurucz93})  \\
H$_2$ quasimolecule absorption                   &  QUASIH  &  
Pavlenko (\cite{pav99a}) \\
C$^-$ ion absorption                             &  CMINUSOP&  
Pavlenko (\cite{pav99a}) \\
Bound-free of C\,{\sc i}                         &  C1OPAC  &  
Pavlenko (\cite{pav99a})  \\ 
\hline
\hline
\end{tabular}
\end{center}
\end{table*}

In the high pressure conditions of the atmospheres of L-dwarfs some 
molecules can be oversaturated (Tsuji et al. \cite{tsuji96a}); in this case
such molecules should undergo condensation. To take into account this 
effect, we reduced the abundances of those molecular species down to the 
equilibrium values (Pavlenko \cite{pav98}). The constants for computations 
of saturation densities were taken from Gurwitz et al. (\cite{gurvitz82}).

We used the set of continuum opacity sources listed in Table~\ref{tab2} 
where we 
also give the original sources for the opacity computation codes. That 
opacity grid allows us to obtain resonable fits to a variety of stars (see 
e.g. Mart\'\i n et al. \cite{martin97b}; Israelian, Garc\'\i a L\'opez \& 
Rebolo \cite{israelian98}; Yakovina \& Pavlenko \cite{yakovina98}). 
Opacities due to molecular band absorption were treated using the Just 
Overlapping Line Approximation (JOLA). Synthetic spectra of late M-dwarfs 
using both the continuum opacities listed in Table~\ref{tab2} and the JOLA 
treatment for molecular band absorptions have been already discussed in 
Pavlenko (\cite{pav97}).

The alkali line data were taken from the VALD database (Piskunov et al. 
\cite{piskunov95}, see their Table~3). At the low temperatures of 
our objects we deal with saturated absorption lines of alkalis. 
Their profiles are pressure broadened. At every depth 
in the model atmosphere the profile of the aborption lines is described 
by a Voigt function $H(a,v)$, where damping constants $a$ were computed 
as in Pavlenko et al. (\cite{pav95}).

\begin{table}
\begin{center}
\caption[]{\label{tab3} List of atomic lines used in our computations}
\begin{tabular}{lccc}
\hline\hline
    & $\lambda_{vac}$ (nm)  &   $gf$  &    $E^{''}$ (eV) \\
\hline
\noalign{\smallskip}
    Li\,{\sc i} & 610.5164 &0.126\,E$+$01 & 1.840  \\
    Li\,{\sc i} & 610.5275 &0.229\,E$+$01 & 1.850  \\
    Li\,{\sc i} & 610.5290 &0.257\,E$+$00 & 1.850  \\
    Li\,{\sc i} & 670.9551 &9.790\,E$-$01 & 0.000  \\
    Li\,{\sc i} & 670.9701 &4.900\,E$-$01 & 0.000  \\
    Li\,{\sc i} & 812.8408 &0.216\,E$+$00 & 1.850  \\
    Li\,{\sc i} & 812.8629 &0.432\,E$+$00 & 1.850  \\
    Na\,{\sc i} & 589.1518 &1.288\,E$+$00 & 0.000  \\
    Na\,{\sc i} & 589.7489 &6.457\,E$-$01 & 0.000  \\
    Na\,{\sc i} & 819.6986 &3.390\,E$-$01 & 2.105  \\
    Na\,{\sc i} & 819.7016 &3.090\,E$+$00 & 2.105  \\
    Na\,{\sc i} & 818.5443 &1.690\,E$+$00 & 2.100  \\
    K\,{\sc i}  & 766.6961 &1.340\,E$+$00 & 0.000  \\
    K\,{\sc i}  & 770.1031 &6.760\,E$-$01 & 0.000  \\
    Rb\,{\sc i} & 780.2348 &1.370\,E$+$00 & 0.000  \\
    Rb\,{\sc i} & 794.9729 &6.800\,E$-$01 & 0.000  \\
    Cs\,{\sc i} & 894.5900 &7.800\,E$-$01 & 0.000   \\
    Cs\,{\sc i} & 852.3285 &1.620\,E$+$00 & 0.000  \\
\hline
\hline
\end{tabular}
\end{center}
\end{table}

\subsection{Molecular opacities}

Detail molecular opacities have been computed for all the molecules listed 
in Table~2 of Pavlenko et al. (\cite{pav95}) as well as for CrH and CaH.
To compute the opacity due to absorption of VO and TiO bands we followed the
scheme presented in Pavlenko et al. (\cite{pav95}) and Pavlenko 
(\cite{pav97}). However, for the $B ^{4}\Pi_{(r)} - X ^{4}\Sigma^{-}$ band 
system of VO we used a more complete matrix of Franc-Condon Factors computed 
by the FRANK program (Cymbal \cite{cymbal77}) with account of 
rotational-vibrational interaction in the Morse-Pekeris approximation 
modified by Schumaker (\cite{schumaker69}, see Pavlenko \cite{pav99b} for 
more details). Futhermore for this band system we used oscillator strength 
$f_e$ from Allard \& Hauschildt (\cite{allard95}). For the $\epsilon$
band system of TiO the parameters of Schwenke (\cite{schwenke98}) were used.  

The CrH molecular band opacity was also considered, the data required 
for the computations of the electronic transition 
A$^{6}\Sigma^{(+)}$--X$^{6}\Sigma^{(+)}$ being taken from Huber \& Herzberg 
(\cite{huber79}). Franc-Condon factors were
computed by the FRANK program (Cymbal \cite{cymbal77}). For this band system 
we used $f_e$ = 0.001, determined by Pavlenko (\cite{pav99b}). Molecular 
bands of the A$^{6}\Sigma^{(+)}$--X$^{6}\Sigma^{(+)}$ of CrH lie in the wide 
wavelength region 500--1200\,nm, and a head of the strong (0,1) band lies 
near the core of the K\,{\sc i} atomic line. A head of the (0,0) band 
of A$^{6}\Sigma^{(+)}$--X$^{6}\Sigma^{(+)}$ system of CrH at 860\,nm is 
blended with the heads of (1,0), (3,2), (2,1) bands of the 
$B ^{4}\Pi_{(r)}$--$X ^{4}\Sigma^{-}$ system of VO.
For the spectral region 650--710\,nm we took into account the absorption 
due to the $B ^2\Sigma$--$X ^2\Sigma$ band system of CaH for which we 
adopted $f_e$\,=\,0.05. Frank-Condon factors were computed with the FRANK 
program. Although FeH is an important absorber at wavelengths around 870\,nm 
and 990\,nm for the coolest dwarfs, it is not yet included in our 
calculations due to the lack of appropriate laboratory data. In the 
wavelength region presented in this paper, the blue band of FeH is 
blended with CrH and VO.

\section{Analysis and interpretation}

\begin{figure}
\begin{center}
\resizebox{7.5cm}{!}{\includegraphics{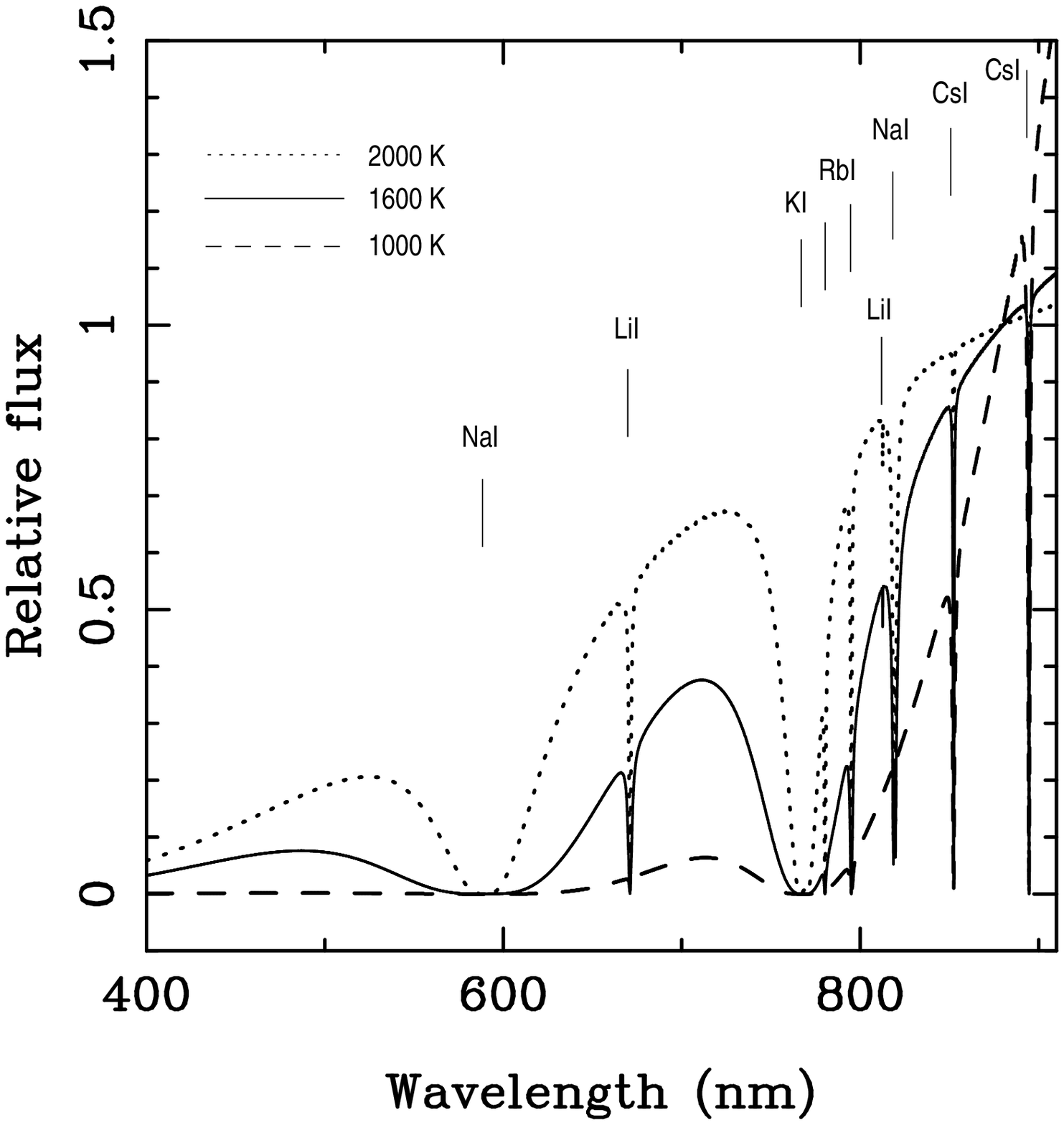}} 
\resizebox{7.5cm}{!}{\includegraphics{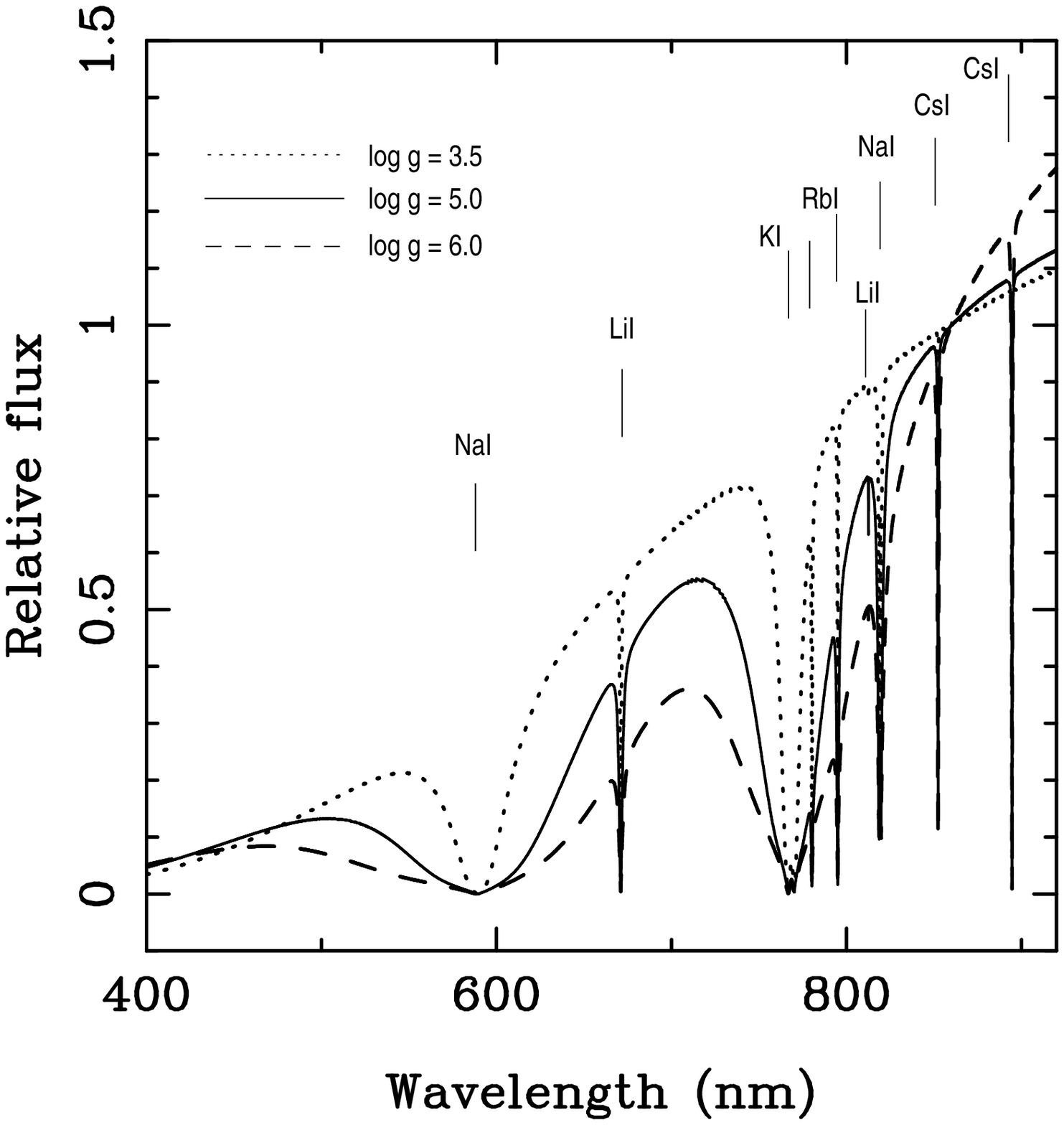}}
\end{center}
\caption[]{\label{fig3} Theoretical optical spectra for the alkali elements 
computed with different values of $T_{\rm eff}$ and gravity. The upper 
panel shows the dependence with temperature (computations performed using 
Tsuji's \cite{tsuji00} C-type models and log\,$g$\,=\,5.0. Spectra 
are normalized at 880\,nm). The lower panel displays the dependence with 
gravity at $T_{\rm eff}$\,=\,1600\,K (Allard's \cite{allard99} ``dusty'' 
models. Spectra are normalized at 860\,nm). Identifications of the atomic 
lines are given in the top.}
\end{figure}

\subsection{Atomic features}

Our spectral synthesis for model atmospheres in the temperature range 
2200--1000\,K confirms the relevant role of atomic features due to the 
Na\,{\sc i} and K\,{\sc i} resonance doublets in the spectral range 
640--930\,nm. The strength of these doublets increases dramatically when 
decreasing the effective temperature ($T_{\rm eff}$). In Fig.~\ref{fig3} 
we display synthetic spectra for different $T_{\rm eff}$ and gravity values. 
These computations do not include any molecular or grain opacity in order to 
show how alkali absorptions change with these parameters. In addition, we can 
see in the figure how the overall shape of the coolest L-dwarf spectra is
governed by the resonance absorptions of Na\,{\sc i} and K\,{\sc i}. 
Even the less abundant alkali (i.e. Li, Rb, Cs) produce lines of remarkable 
strength. Note the increase in intensity of K\,{\sc i} and Na\,{\sc i} lines
with decreasing $T_{\rm eff}$ and with increasing atmospheric gravity.
Although this different behaviour can, in principle, make it difficult
to disentangle these parameters from the optical spectra, simple physical
considerations give an upper limit to gravity of log\,$g$\,=\,5.5 for 
brown dwarfs with lithium and therefore the large broadening of the 
K\,{\sc i} lines seen in some of our objects cannot be attributed to higher 
aphysical values of gravity.

Our computations provide a qualitative explanation of the far-red spectral 
energy distributions presented in Fig.~\ref{fig1}. The equivalent widths 
(EWs) of the K\,{\sc i} and Na\,{\sc i} resonance doublets may reach several 
thousand Angstroms, becoming  the strongest lines so far seen in the spectra 
of astronomical objects. This is mainly caused by the high pressure
broadening in the atmospheres of the coolest dwarfs, where damping constants 
of the absorption lines vary from 0.001 in the uppermost layer of the 
atmosphere to 2--4 in the deepest regions. The subordinate lines of 
Na\,{\sc i} at 819.5\,nm, clearly seen in all the L-dwarfs in our sample, 
become weaker as $T_{\rm eff}$ decreases. These computations also show that 
the subordinate Li\,{\sc i} line at 812.6\,nm may be detected in early/mid 
L-dwarfs with equivalent widths not exceeding 1\AA, and that the triplet 
at 610.3\,nm appears completely embedded by the wings of the K\,{\sc i} 
and Na\,{\sc i} resonance lines.

\begin{figure}
\begin{center}
\resizebox{7.5cm}{!}{\includegraphics{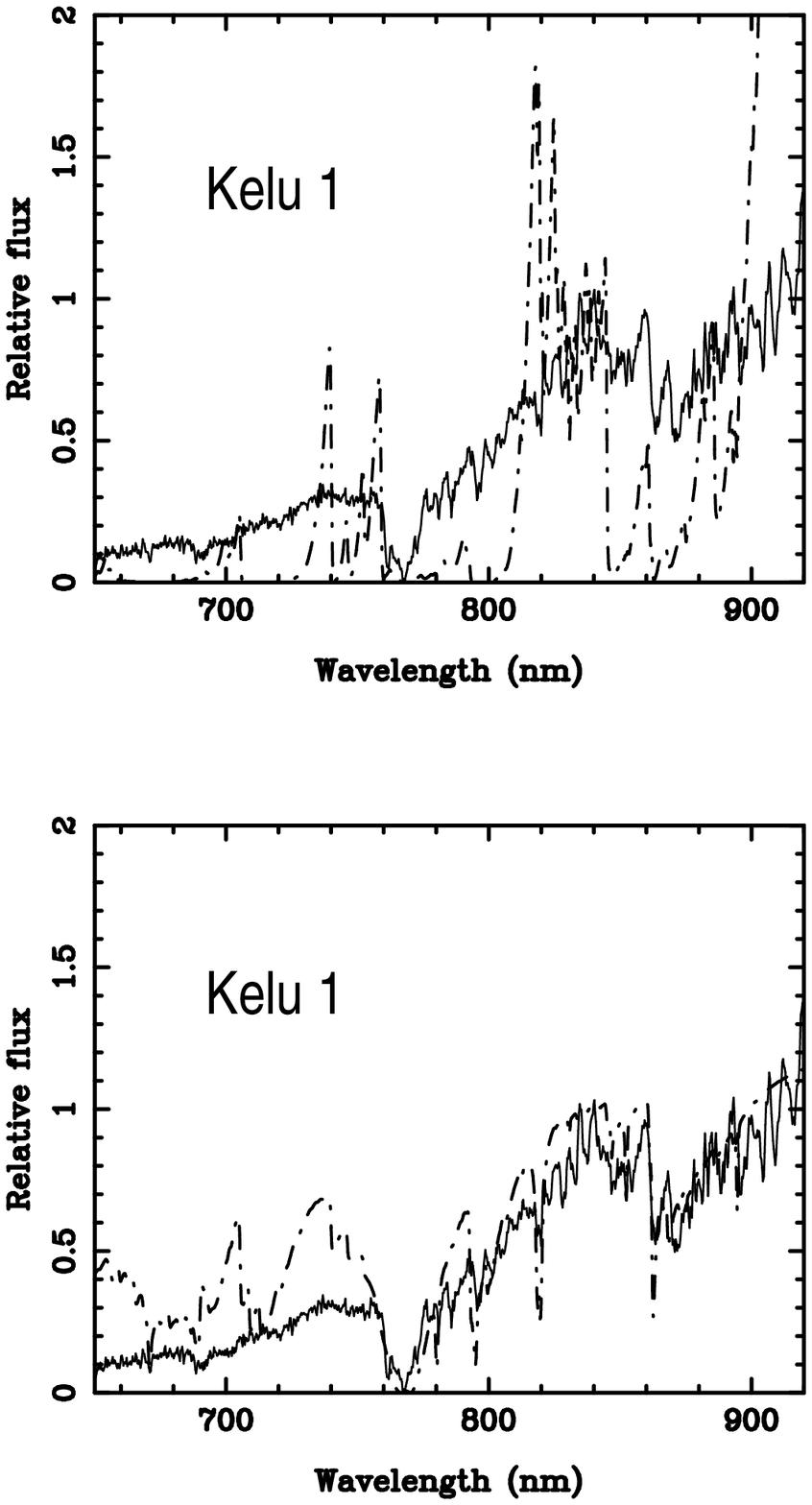}}
\end{center}
\caption[]{\label{fig4} The observed spectrum of Kelu\,1 normalized at 
$\sim$840\,nm is shown by the full line and the predicted spectra 
(Tsuji's C-type model atmosphere for $T_{\rm eff}$\,=\,2000\,K and 
log\,$g$\,=\,5) are shown by the dash-dotted line at a resolution of 
10\AA. Upper panel displays 
the theoretical spectrum computed considering only the natural depletion 
of molecules as a result of chemical equilibrium. Lower panel depicts 
the same computations taking into account an ``extra'' depletion of TiO 
($R$\,=\,0.01), VO ($R$\,=\,0.08), CaH ($R$\,=\,0.5) and CrH ($R$\,=\,0.2).}
\end{figure}

\begin{figure}
\begin{center}
\resizebox{7.5cm}{!}{\includegraphics{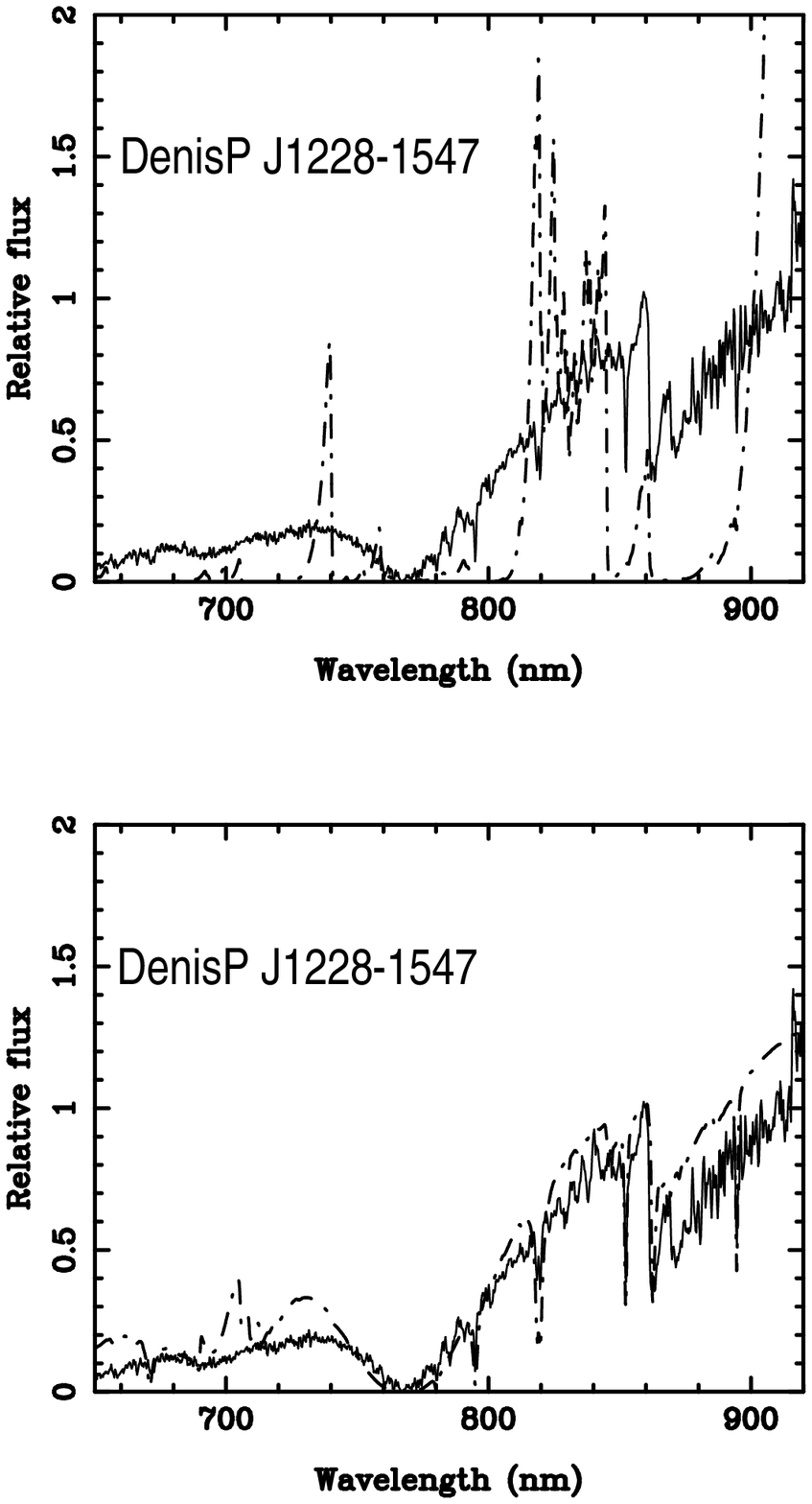}}
\end{center}
\caption[]{\label{fig5}The observed spectrum of DenisP\,J1228--1547 
normalized at $\sim$860\,nm is shown by the full line and the predicted 
spectra (Tsuji's C-type model atmosphere for $T_{\rm eff}$\,=\,1600\,K and 
log\,$g$\,=\,5) are shown by the dash-dotted line at a resolution of 
10\AA. Upper panel displays 
the theoretical spectrum computed considering only the natural depletion 
of molecules as a result of chemical equilibrium. Lower panel depicts 
the same computations taking into account an ``extra'' depletion of TiO 
($R$\,=\,0.0), VO ($R$\,=\,0.0), CaH ($R$\,=\,0.5) and CrH ($R$\,=\,0.04).}
\end{figure}

\subsection{Molecular features}

In the optical spectra of very cool dwarfs one expects the presence of 
bands of VO, TiO and indeed, our synthetical spectra show that these 
bands play an important role in the cases of BRI\,0021--0214 and Kelu\,1. 
In Fig.~\ref{fig4} (upper panel) we compare the observed spectrum of Kelu\,1 
with a synthetic spectrum obtained using the Tsuji C-type model for 
$T_{\rm eff}$\,=\,2000\,K and log\,$g$\,=\,5. The natural depletion of 
molecules resulting from the chemical equilibrium is not sufficient to get a 
reasonable fit to the data. The discrepancies with respect to the observed 
spectrum can be notably reduced (Fig.~\ref{fig4}, lower panel) if we impose 
a depletion of CaH, CrH, TiO and VO molecules which should account for 
the condensation of Ca, Cr, Ti and V atoms into dust grains. Since we do 
not have an appropriate theoretical description of the processes of dust 
formation at present, we use the simple approach of modifying the chemical 
equilibrium for all molecules. We implemented this ``extra'' depletion 
simply by introducing a factor $R$ which describes the reduction of 
molecular densities of the relevant species over the whole atmosphere. 
From the comparison between observed and computed spectra we find that the 
reduction factor for TiO ranges from 0 (complete depletion) to 1 
(i.e. non depletion). In Fig.~\ref{fig5} we can see the comparison of
similar synthetic spectra (Tsuji's C-type model atmosphere for 
$T_{\rm eff}$\,=\,1600\,K and log\,$g$\,=\,5) with the spectrum of 
DenisP\,J1228-1547. Total or almost total depletion of Ti and V into the 
dust grains is required to explain the spectrum of the mid-type L-dwarfs 
DenisP\,J1228--1547 and DenisP\,J0205--1159.

\begin{figure}
\begin{center}
\resizebox{7.5cm}{!}{\includegraphics{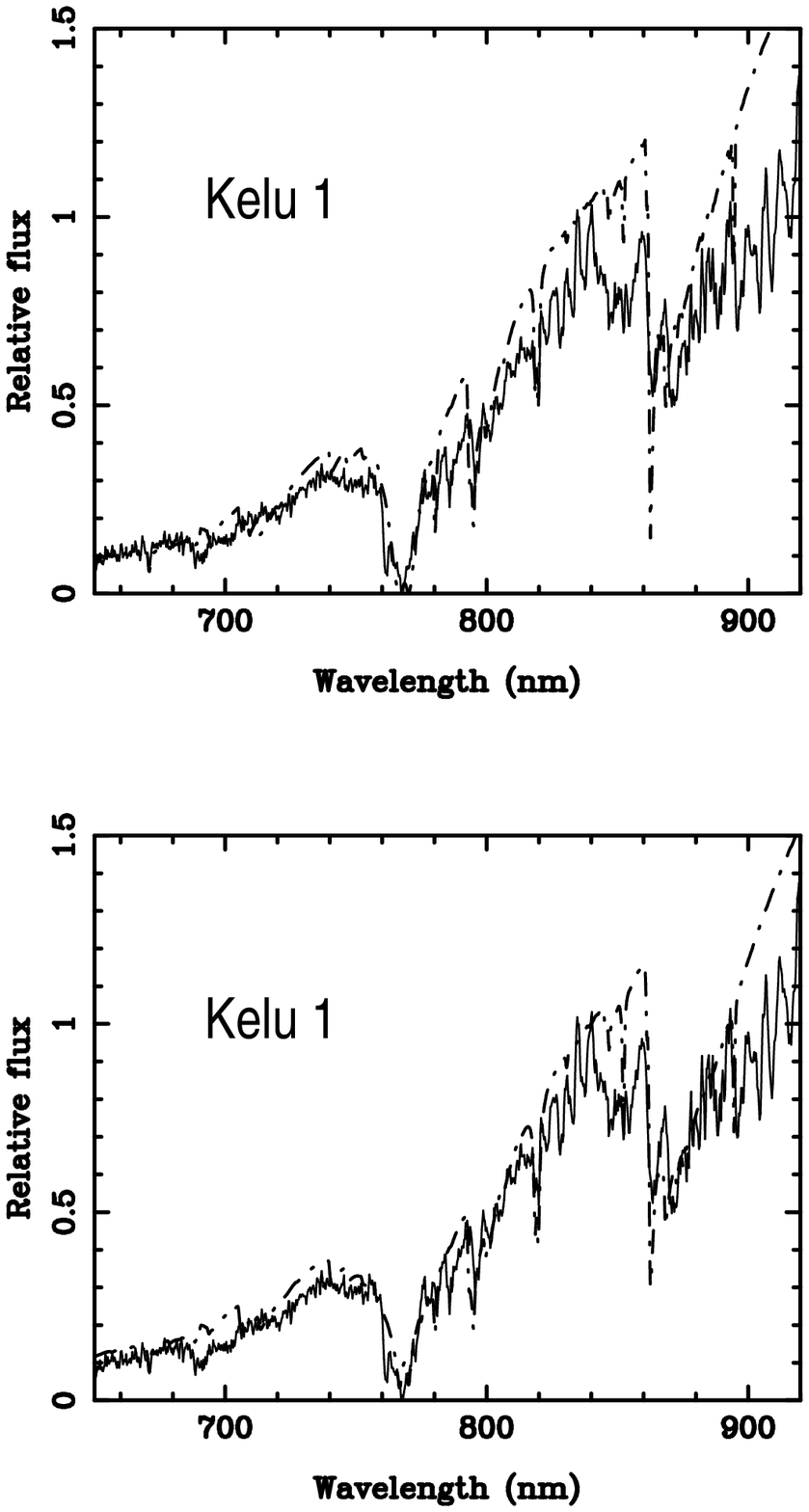}}
\end{center}
\caption[]{\label{fig6} Kelu\,1's observed spectrum (full line) compared 
to the best fits (dash-dotted line) obtained using Tsuji's (\cite{tsuji00}) 
C-type models (upper panel, 
$T_{\rm eff}$\,=\,2000\,K and log\,$g$\,=\,5), and Allard's (\cite{allard99}) 
dusty models (lower panel, $T_{\rm eff}$\,=\,1800\,K and log\,$g$\,=\,5). 
Both predicted spectra with a resolution of 10\AA~have been computed 
considering the $R$ depletion factors given in Fig.~\ref{fig4} and the 
AdO law described in the text with $a_{\circ}$\,=\,0.03.}
\end{figure}

\begin{figure}
\begin{center}
\resizebox{7.5cm}{!}{\includegraphics{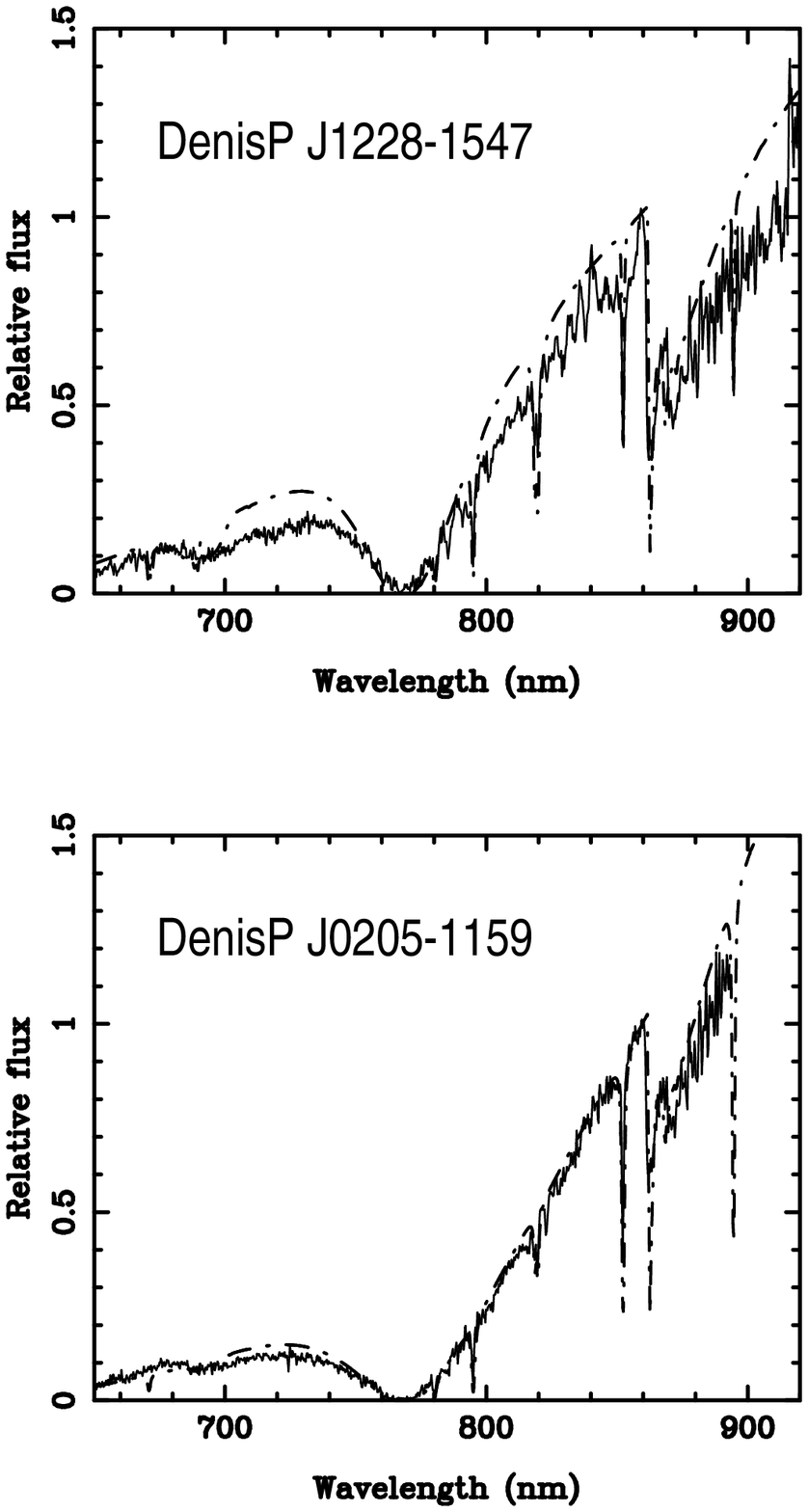}}
\end{center}
\caption[]{\label{fig7} DenisP\,J1228--1547 and DenisP\,J0205--1159's 
observed data (full line) compared to the best fits (dash-dotted line) 
obtained using Tsuji's 
(\cite{tsuji00}) C-type models (upper panel: $T_{\rm eff}$\,=\,1600\,K and 
log\,$g$\,=\,5; lower panel: $T_{\rm eff}$\,=\,1200\,K and log\,$g$\,=\,5). 
Both predicted spectra with a resolution of 10\AA~have been computed 
considering the $R$ depletion factors given in Fig.~\ref{fig5} and the 
AdO law described in the text with $a_{\circ}$\,=\,0.006.}
\end{figure}

\subsection{The need for additional opacity (AdO) \label{sec_ado}}

Our first attempts to model the spectra including atomic and molecular
features showed in Figs.~\ref{fig4} and~\ref{fig5} were only modestly 
successful. Although we could reproduce reasonably well the red wing of 
the K\,{\sc i} line, the theoretical fluxes in the blue part of
our synthetic spectra (640--750\,nm) are too large. In fact, we cannot 
fit the observations by taking into account only the opacity provided 
by the Na\,{\sc i} and K\,{\sc i} resonance doublets and the continuum 
opacity sources listed in Table~\ref{tab2}. 

Since the formation of dust in these cool atmospheres can produce
additional opacity (AdO) which may affect the synthetic spectra we decided
to investigate whether a simple description of it 
could help us to improve the comparison between observed and computed
spectra. We adopted as law for AdO the following 
expression: $a_{\nu}=a_{o}*(\nu/\nu_{0})^N$. For $N$\,=\,0 to
4 this law corresponds to the case of radiation scattering produced by
particles of different sizes, being $N$\,=\,4 the case of pure Rayleigh
scattering, and $N$\,=\,0 corresponding to the case of white scattering. 
However, at present we cannot distinguish whether this AdO is due to 
absorption or scattering processes. The parameters $N$ and $a_{0}$ would 
be determined from the comparison with observations, but in all cases we 
try to get the best fit for $N$\,=\,4, which would be the most simple
from the physical point of view. We adopted as $\nu_o$ the frequency 
of the K\,{\sc i} resonance line at 769.9\,nm. 
Our model of AdO is depth independent and therefore in this
approach we cannot model the inhomogeneities (e.g. dust clouds)
which may exist in L-dwarf atmospheres.

\begin{figure}
\begin{center}
\resizebox{7.5cm}{!}{\includegraphics{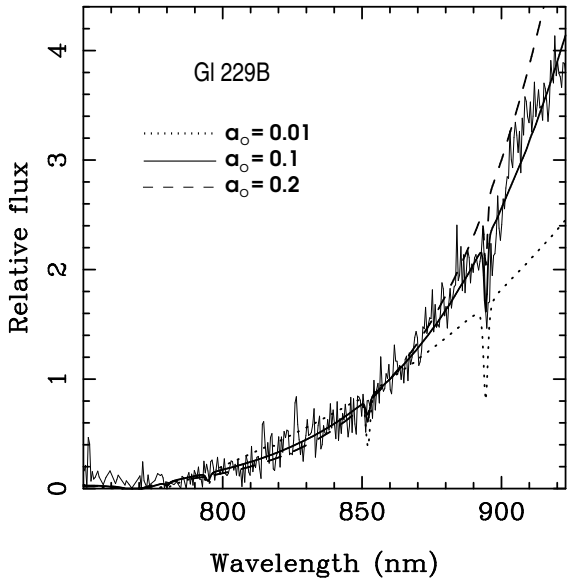}}
\end{center}
\caption[]{\label{fig8} Gl\,229B observed data (thin full line) compared to 
some predicted spectra obtained using the Tsuji's (\cite{tsuji00}) C-type 
model of $T_{\rm eff}$\,=\,1000\,K and log\,$g$\,=\,5. Different values of 
$a_{\circ}$ are oveplotted to show the effect of dust opacity at this 
low temperature. The best fit is provided by $a_{\circ}$\,=\,0.1 and 
$N$\,=\,4. Only Cs\,{\sc i} features are seen in the wavelenth range 
presented here.}
\end{figure}

We have investigated how our simple approach for the modelling of AdO may 
lead to better comparisons between predicted and observed spectra. The 
implementation of our law of AdO depresses the fluxes in the blue wing 
of the K\,{\sc i} doublet, considerably improving the reproduction of 
the observed data of Kelu\,1 (Fig.~\ref{fig6}), DenisP\,J1228--1547 
(Fig.~\ref{fig7}, upper panel) and DenisP\,J0205--1159 (Fig.~\ref{fig7}, 
lower panel). For each object, synthetic spectra were computed for a range 
of $T_{\rm eff}$, gravities, $a_0$ and depletion factors $R$ for TiO, VO, 
CaH and CrH. The previous figures display those syntheses which better 
reproduce the observed spectra. 
From our computations we infer that VO is less 
efficiently depleted than TiO at a given $T_{\rm eff}$. The depletion
of these oxides appears to increase very rapidly as we go from 
BRI\,0021--0214 and Kelu\,1 to the lower temperature objects 
DenisP\,J0205--1159 and DenisP\,J0205--1159. Note the good fit to the 
observed spectra at 860\,nm provided by the (0,0) band of the 
A$^{6}\Sigma^{(+)}$--X$^{6}\Sigma^{(+)}$ system of CrH. From the study of 
this band we also find that the depletion factor of CrH increases from 
the warmer to the coolest objects in the sample. Finally, we also find 
that atomic lines become weaker and narrower when increasing the amount 
of AdO in the atmospheres.

We have also studied whether we can explain the optical spectrum of 
Gl\,229B using the following Tsuji's model: C-type, 
$T_{\rm eff}$\,=\,1000\,K, and log\,$g$\,=5.0, and the AdO law of index 
$N$\,=\,4 used above. In Fig.~\ref{fig8} we show several spectral synthesis
reproducing the optical spectrum of this object, and showing the effect 
of different $a_\circ$ parameters which is related to the amount of dust 
in the atmosphere. In Gl\,229B we need the highest value of $a_0$, which 
is interpreted as evidence for the most ``dusty'' atmosphere in our sample.
We have not attempted to reproduce the Cs\,{\sc i} lines in Gl\,229B. 
According to our hypothesis, the inclusion of AdO 
avoids the contribution of high pressure regions to the formation of
these lines. The shorter wavelength Cs\,{\sc i} line at 894.3\,nm is very  
affected by dust opacity.

\section{Discussion}

\subsection{$T_{\rm eff}$ for L-type dwarfs}

\begin{table*}
\caption[]{\label{tab4} $T_{\rm eff}$ estimations for our sample adopting 
Tsuji's (\cite{tsuji00}) C-type models and log\,$g$\,=\,5.0.}
\begin{center}
\begin{tabular}{llcccl}
\hline\hline
Object              &  Sp. Type & $I-J$ & $T_{\rm eff}$ & Other measures  & Source\\
                    &           &       & ($\pm$200\,K) & \\
\hline
BRI\,0021--0214     &  M9.5     & 3.30  & 2200          & 1980, 2300      & TMR93, LAH98 \\
Kelu\,1             &  L2 (L2)  & 3.50  & 2000          & 2000, 1900      & B99, RLA97\\
DenisP\,J1228--1547 &  L4.5 (L5)& 3.81  & 1600          & 1800            & B99 \\
DenisP\,J0205--1159 &  L5 (L7)  & 3.82  & 1200          & 1700, 1800      & B99, TK99\\ 
\hline
\end{tabular}
\end{center}
NOTES. Spectral types are given in Mart\'\i n et al. (\cite{martin99}). 
Those spectral types in brackets come from Kirkpatrick et al. 
(\cite{kirk99a}).\\
$(I-J)$ colors have been taken from Leggett, Allard \& Hauschildt 
(\cite{leggett98}).\\
References: TMR93 = Tinney, Mould \& Reid (\cite{tinney93}); 
LAH98 = Leggett et al. (\cite{leggett98}); 
B99 = Basri et al. (\cite{basri99}); RLA97 = Ruiz et al. 
(\cite{ruiz97}); TK99 = Tokunaga \& Kobayashi (\cite{tokunaga99}).
\end{table*}

Our computations provide a reasonable description of the far-red optical 
spectra of L-dwarfs and provide a physical basis for a progressively 
decreasing $T_{\rm eff}$ for the proposed spectral classifications 
(Mart\'\i n et al. \cite{martin99}; Kirkpatrick et al. \cite{kirk99a}). 
Effective temperatures for a few field L-dwarfs and for Gl\,229B have been 
derived from spectra at IR wavelengths (Allard et al. \cite{allard96}; 
Marley et al. \cite{marley96}; Matthews et al. \cite{matthews96}; Tsuji 
et al. \cite{tsuji99}; Jones et al. \cite{jones96}; Kirkpatrick et al. 
\cite{kirk99b}). Here we study to what extent we can use the broad energy 
spectral distribution in the optical to infer the $T_{\rm eff}$ for the 
cool dwarfs in our sample. Our best estimates using Tsuji's models (see 
Table~\ref{tab4}) are in good agreement with those found from IR data for 
objects of similar spectral types and for Gl\,229B. Using Allard's models 
and the simple approach of AdO described in section~\ref{sec_ado} we are 
also able to reproduce the observed spectra, albeit we require lower 
$T_{\rm eff}$'s by up to several hundred degrees for the coolest L-dwarfs. 
This is mainly due to the hotter stratification of Allard's model 
atmospheres for the potassium and sodium lines forming layers 
(see Fig.~\ref{fig2}).
Opposite to IR-based temperature determinations, the estimation from 
optical spectra is very sensitive to the input physics parameters, like 
dust formation, molecular equilibrium, sources of opacity and atmospheric 
models, which limits the accuracy of the estimates to $\sim$200\,K. On the 
other hand, the optical spectra provide an opportunity to test the 
reliability of the physical description of the atmospheres.

Cs\,{\sc i} lines in the optical spectra have been recently used to 
infer the $T_{\rm eff}$ of some L-dwarfs (Basri et al. \cite{basri99}).
In spite of the sensitivity of these lines to many input parameters in 
the models (i.e. to the amount of opacity, chemical equilibrium, etc.) 
we find in general a good agreement with their estimated temperatures for 
the earlier L-dwarfs (see Table~\ref{tab4}). However, for the latest 
L-type objects in our sample, DenisP\,J1228--1547 and DenisP\,J0205-1159, 
we find temperatures up to 400\,K lower. We may attribute this discrepancy 
to the effects that dust opacity has in the formation of optical
lines of alkalie. This produces a kind of ``veiling"  of the atomic lines, 
reducing their intensities. Basri et al. (\cite{basri99}) did not consider
this effect, and therefore they required hotter models in order to explain 
the strength of the observed Cs lines. Another possible reason for the 
discrepancy is that the temperatures in Table~\ref{tab4} have been 
obtained for gravity log\,$g$\,=\,5; if we increased gravity by 0.5\,dex 
we would have to increase the temperatures of our models by about 200\,K. 
In this case the spectral synthesis does not reproduce the observations 
so well but they are still acceptable.

\begin{figure}
\begin{center}
\resizebox{7.5cm}{!}{\includegraphics{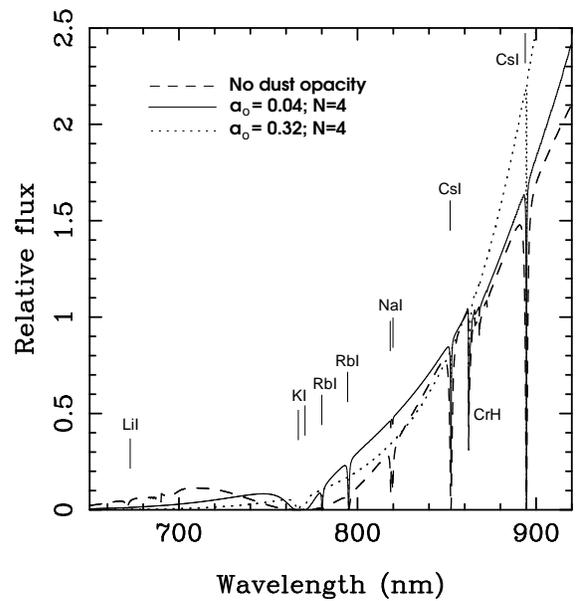}}
\end{center}
\caption[]{\label{fig9} Predicted optical spectra (Tsuji's C-type model 
atmosphere, $T_{\rm eff}$\,=\,1200\,K and log\,$g$\,=\,5) for intermediate 
objects between Gl\,229B and DenisP\,J0205--1159. Different intensities 
in the dust opacity have been considered. Spectra have been normalized 
at 860\,nm. Identification of the atomic 
and molecular bands included in the computations are also provided.}
\end{figure}

\begin{figure}
\begin{center}
\resizebox{7.5cm}{!}{\includegraphics{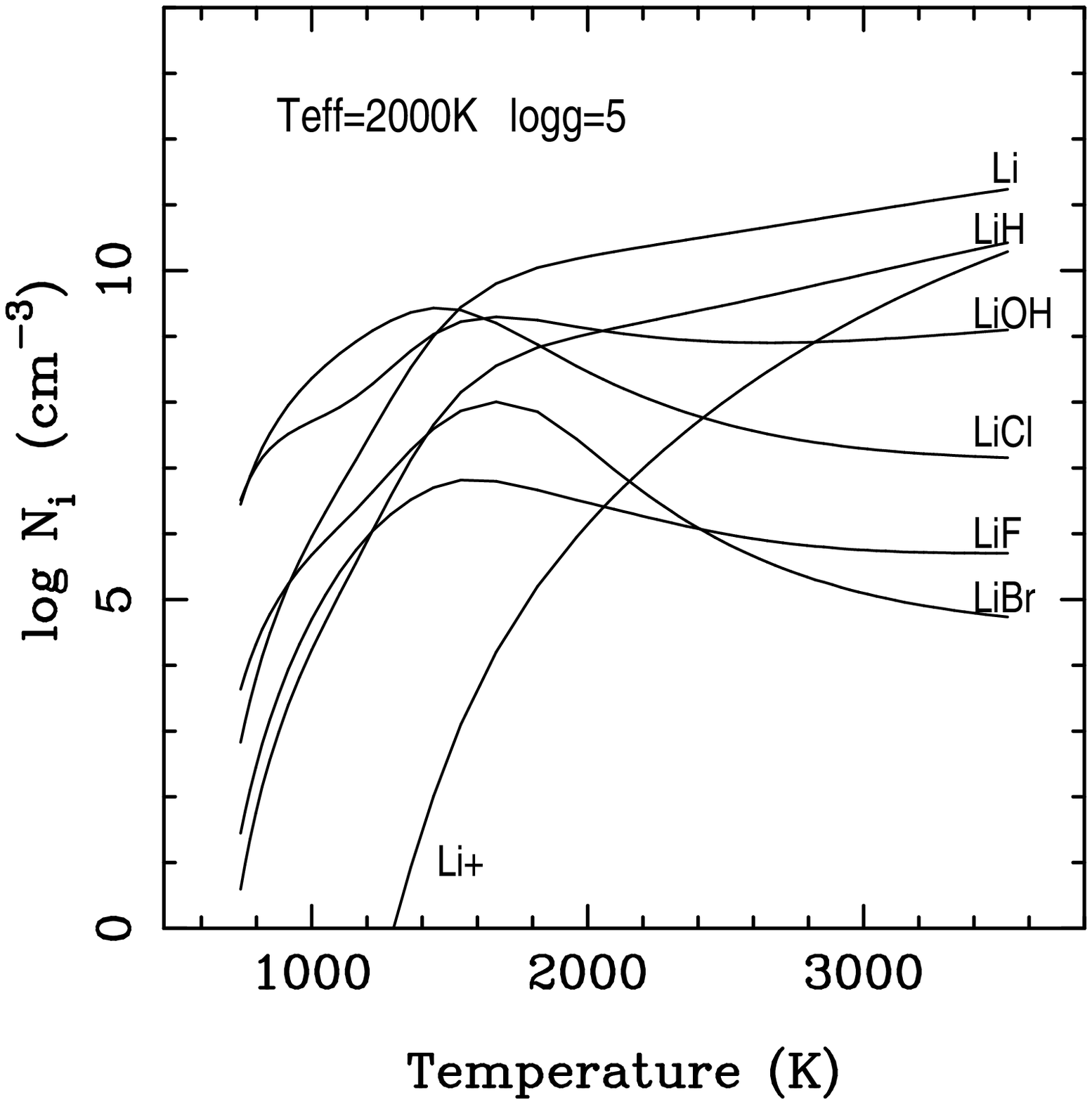}}
\resizebox{7.5cm}{!}{\includegraphics{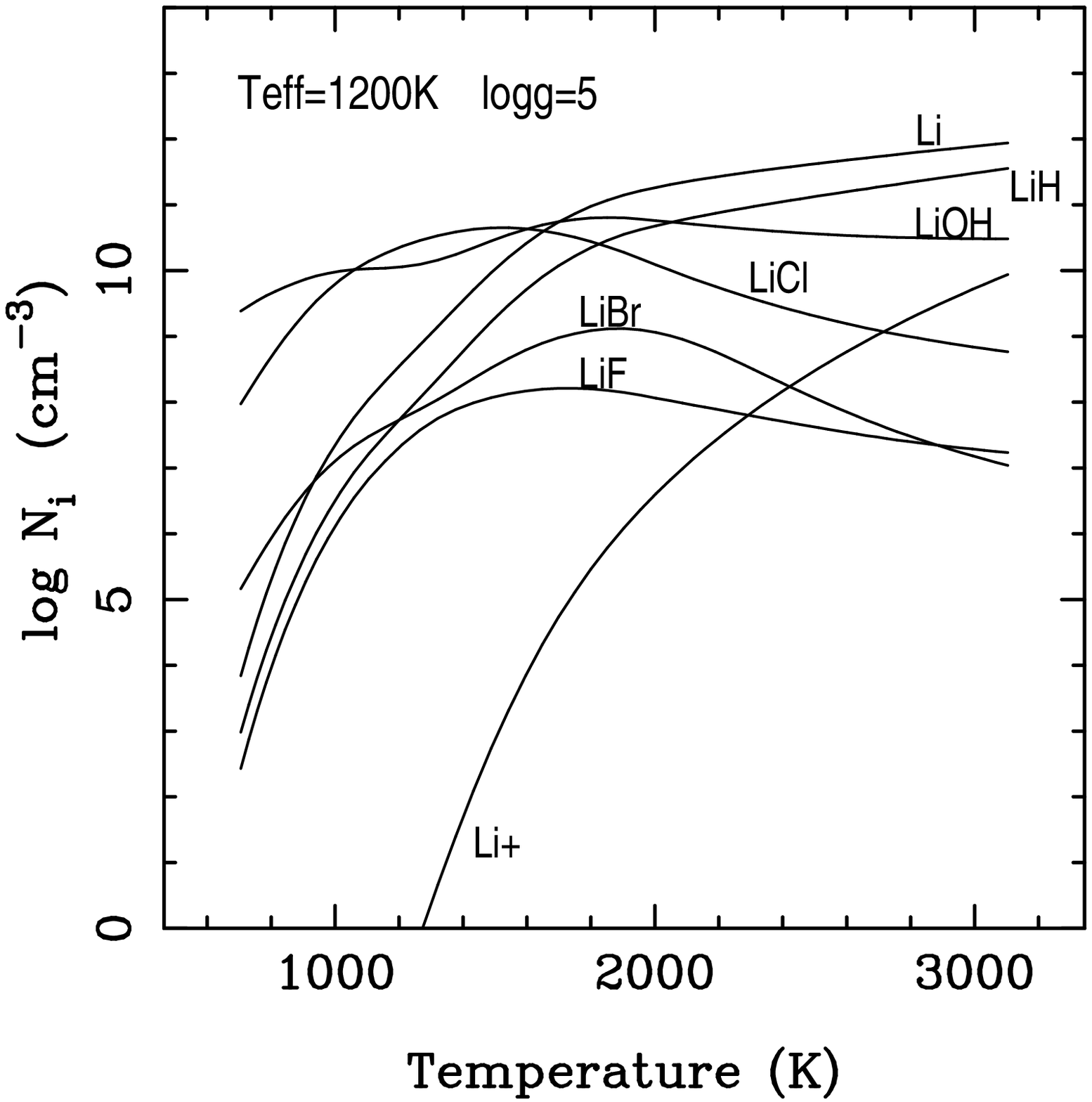}}
\end{center}
\caption[]{\label{fig10} Densities of lithium and the most abundant species containing lithium are plotted for the C-type Tsuji's (\cite{tsuji00}) model atmospheres with $T_{\rm eff}$\,=\,2000\,K (upper panel) and 1200\,K (lower panel) and log\,$g$\,=\,5. Computations have been performed under the assumption of complete chemical equilibrium for a solar elemental mixture.}
\end{figure}

\begin{figure}
\begin{center}
\resizebox{7.5cm}{!}{\includegraphics{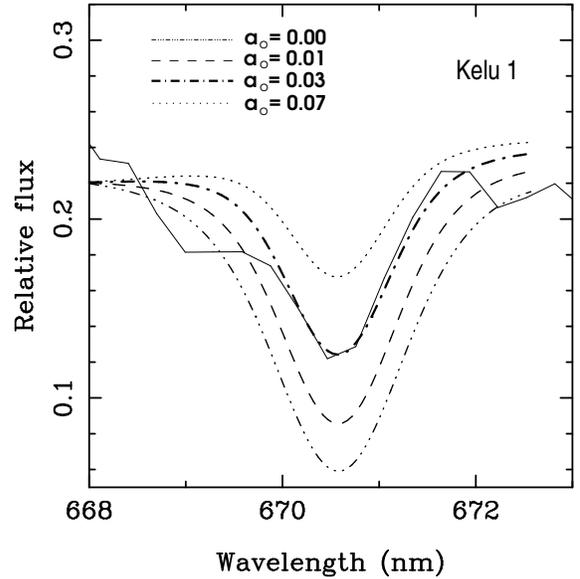}}
\end{center}
\caption[]{\label{fig11} Fitting of the Li\,{\sc i} resonace line of 
Kelu\,1 (full line) using the Tsuji's C-type model atmosphere for 
$T_{\rm eff}$\,=\,2000\,K, log\,$g$\,=\,5. Computations have been performed 
for a lithium abundace of log\,$N$(Li)\,=\,3.0 and considering different 
amounts of dust opacity. The resolution of all theoretical spectra is the 
same than the one of the observed data. The predicted model that better 
fits the observations coincides with the one that also nicely reproduces 
the overall shape of the optical spectrum ($a_\circ$\,=\,0.03, see 
Fig.~\ref{fig6}).}
\end{figure}

\subsection{Alkali lines: the case of lithium}

Recently, several new Gl\,229B-like objects have been discovered by the 
SDSS survey (Strauss et al. \cite{strauss99}) and the 2MASS collaboration 
(Burgasser et al. \cite{burgasser99}). Based on their near-IR spectra, 
these authors suggest that these new cool brown dwarfs may be warmer than 
Gl\,229B. In order to estimate the optical properties of these objects we 
have computed synthetical spectra using a Tsuji's C-type model of 
$T_{\rm eff}$\,=\,1200\,K and log\,$g$\,=\,5.0, and we have adopted the 
basic prescriptions that were followed in section~4, i.e total depletion 
of VO and TiO and the AdO law with $N$\,=\,4. In Fig.~\ref{fig9} we plot 
the resulting spectra considering different amounts of dust opacity. 
As expected, the overall shape of the spectrum is intermediate between 
that of DenisP\,J0205--1159 and Gl\,229B. In the absence of dust absorption 
($a_{\circ}$\,=\,0.0) alkali lines are clearly seen (including the lithium 
resonance doublet), and the spectrum is governed by the sodium and 
potassium lines. If we consider dust opacities comparable to those in 
Gl\,229B, the alkali lines of Cs and Rb become weaker, but still 
detectable with intermediate resolution spectroscopy. Remarkably, 
the subordinate Na\,{\sc i} doublet at 819.5\,nm is very sensitive to 
the incorporation of dust opacity due to the larger depths of its 
formation as compared with resonance lines. For very high dust opacities 
these lines may become undetectable. 

The effects of additional dust opacity on the formation of Li\,{\sc i} 
lines (resonance and subordinate ones) also deserve detailed consideration 
since they play a major role as a discriminator of substellar nature for 
brown dwarf candidates (see Rebolo, Mart\'\i n, \& Magazz\`u \cite{rebolo92}; 
Magazz\`u, Mart\'\i n, \& Rebolo \cite{magazzu93}). Most of the known brown 
dwarfs are actually recognized by the detection of the Li\,{\sc i} 
resonance doublet in their spectra (Rebolo et al. \cite{rebolo96}; 
Mart\'\i n et al. \cite{martin97a}; Rebolo et al. \cite{rebolo98}; 
Tinney \cite{tinney98}; Kirkpatrick et al. \cite{kirk99a}). 
The chemical equilibrium of lithium contained molecules have been 
considered in all our syntheses (Fig.~\ref{fig10} depicts the density 
profiles for lithium species for two C-type models by Tsuji \cite{tsuji00}).
Our computations show that both, the resonance line at 670.8\,nm, and the 
subordinate lines at  601.3\,nm and 812.6\,nm are very sensitive to the 
AdO that we need to incorporate in the spectral synthesis if we want to 
explain the observed broad spectral energy distribution. Among the 
subordinate lines the doublet at 812.6\,nm is more easily detectable, but 
the predicted EWs, assuming fully preserved lithium, are rather small, 
ranging from EW\,=\,0.4\AA~to 0.04\AA~for $T_{\rm eff}$ values in the range 
2000\,K down to 1200\,K. These EWs are considerably reduced by the 
inclusion of the AdO  described in the previous section, which makes their 
detection rather difficult. In Table~\ref{tab5} we give the predicted 
EWs of the Li\,{\sc i} resonance doublet at 670.8\,nm for several of the 
coolest model atmospheres (2000--1000\,K) considered in this work.
First, we note that in the absence of any AdO (second column in the table), 
we would expect rather strong neutral Li resonance lines in the spectra 
of objects 
as cool as DenisP\,J0205--1159 and Gl\,229B. The chemical equilibrum of 
Li-contained species still allow a sufficient number of Li atoms to produce 
a rather strong resonance feature; one reason for this is that Cl and O 
atoms should be also bounded into other molecules (e.g. NaCl, KCl, H$_{2}$O, 
etc.). Our computations indicate that objects like DenisP\,J0205--1159 and 
cooler objects with moderate dust opacities should show the Li\,{\sc i} 
resonance doublet if they had preserved this element from nuclear burning, 
and consequently the lithium test can still be applied. Furthermore, even 
in very dusty cool atmospheres like that of Gl\,229B for which we have 
inferred a high value of the opacity parameter of $a_\circ$\,=\,0.1 
(fourth column in Table~\ref{tab5}), the lithium resonance line could 
be detected with an EW of several hundred m\AA ~ (high S/N data 
would be required). 

Another effect that we shall consider is whether small changes in the AdO 
(which could be originated as a consequence of some ``meteorological'' 
phenomena occurring in these cool atmospheres) can lead to detectable 
variations in the EWs of the lithium lines. In particular, weak lithium 
lines do not necessarily imply a depletion of this element. The observed 
Li\,{\sc i} variability in Kelu\,1 (with changes in EW by a factor 5),
could be an indication of meteorological changes in the atmosphere
of this rapidly rotating cool object. In Fig.~\ref{fig11} we present 
several spectral synthesis showing the sensitivity of the lithium line 
to the AdO in the atmosphere. The AdO parameters which give 
the best fit for the lithium line in Kelu\,1 (EW\,=\,6.5$\pm$1.0\,\AA) 
coincide with those also providing the best fit to the whole optical 
spectrum (see Fig.~\ref{fig6}). Anyway, the obtained lithium abundance 
is consistent with complete preservation of this element.

\begin{table}
\caption[]{\label{tab5} Equivalent widths (\AA) of the Li\,{\sc i} resonance 
doublet at 670.8\,nm computed for the C-type Tsuji's (\cite{tsuji00}) model 
atmospheres, cosmic Li abundance 
(log\,$N$(Li)\,=\,3.2) and gravity log\,$g$\,=\,5.0.}
\begin{center}
\begin{tabular}{crrr}
\hline\hline
\multicolumn{1}{c}{} &
\multicolumn{3}{c}{$a_{\circ}$}   \\
\multicolumn{1}{c}{$T_{\rm eff}$} &
\multicolumn{3}{c}{\rule{2.5cm}{0.1mm}}   \\
\multicolumn{1}{c}{} & 
\multicolumn{1}{c}{0.00}  & 
\multicolumn{1}{c}{0.01}  & 
\multicolumn{1}{c}{0.10}  \\
\multicolumn{1}{c}{} &
\multicolumn{3}{c}{\rule{2.5cm}{0.1mm}}   \\
\multicolumn{1}{c}{(K)} &
\multicolumn{3}{c}{EW (\AA)}   \\
\hline
  1000 & 17 &  8 & 0.6  \\
  1200 & 30 & 12 & 0.7  \\
  1400 & 42 & 21 & 0.9  \\ 
  1600 & 40 & 24 & 1.6  \\
  2000 & 23 & 16 & 3.6  \\
\hline
\end{tabular}
\end{center}
\end{table}

\section{Conclusions}

In this paper we have attempted to model the far-red spectra (640--930\,nm) 
of several L-dwarfs suitably selected to cover this new spectral class. 
We have used model atmospheres from Tsuji (\cite{tsuji00}) and Allard 
(\cite{allard99}), as well as an LTE spectral synthesis code 
(Pavlenko et al. \cite{pav95}) which takes into account chemical 
equilibrium for more than 100 molecular species, and detailed opacities 
for the most relevant bands. We have arrived to the following conclusions:

1) Alkali lines play a major role governing the far-red spectra of L-dwarfs.
At early types, this role is shared with TiO and VO bands, which dominate 
this spectral region in late M-dwarfs. As we move to later spectral types 
we need to incorporate progressively higher depletions of these oxides 
and of the hydrides CrH and CaH, 
consistently with the expectation that Ti and V atoms are depleted into 
grains; and we also require additional opacity to reproduce the overall 
shape of the spectra. This additional opacity could be either due to 
molecular/dust absorption or to dust scattering.

2) We have shown that a simple law for this additional opacity of the form 
$a_{\circ} \ (\nu /\nu_{\circ})^N$, with $N$\,=\,4, gives a sufficiently 
good fit to the observed spectra of L-dwarfs and Gl\,229B. For this late 
object we require the highest value of $a_{\circ}$ consistent with a 
very dusty atmosphere. The strength of alkali lines is highly affected 
by this opacity.

3) From the best fits to our spectra, we derive the most likely 
$T_{\rm eff}$ values for our sample of L-dwarfs. For the warmer objects, 
our values are consistent with  those obtained by other authors, however 
we find lower $T_{\rm eff}$'s by serveral hundred degrees for the coolest
L-dwarfs. Because the optical spectra are very much affected by the input 
physics, a more reliable $T_{\rm eff}$ scale should be obtained by fitting 
the IR data of these cool objects.

4) After detailed consideration of chemical equilibrium, we find that the 
lithium resonance doublet at 670.8\,nm can be detected in the whole 
spectral range (down to 1000\,K). In the coolest L-dwarfs the strength of 
the resonance line is more affected by the amount of additional opacity 
needed to explain the spectra than by the depletion of neutral lithium 
atoms into molecular species. In those atmospheres where the additional 
opacity required is low, the lithium test can provide a useful 
discrimination of substellar nature. Changes 
in the physical conditions governing dust formation in L-dwarfs, will 
cause variability of the lithium resonance doublet. Taking into account 
the need for additional opacity in Kelu\,1, we find that the lithium 
abundance can be as high as log\,$N$(Li)\,=\,3.0, i.e. consistent with 
complete preservation.

\begin{acknowledgements}
We thank T. Tsuji, F. Allard, D. Schwenke, G. Schultz and B. Oppenheimer 
for providing us model atmospheres, updated TiO molecular data and the 
optical spectrum of GL\,229B, respectively. We are also indebted to 
R. Garc\'\i a L\'opez, Gibor Basri and Eduardo L. Mart\'\i n for their 
assistance with the observations.
Partial financial support was provided by the Spanish DGES project no.
PB95-1132-C02-01.

\end{acknowledgements}


\begin{thebibliography}{}

\bibitem[1999]{allard99} Allard, F. 1999, private communication

\bibitem[1995]{allard95} Allard, F., Hauschildt, P. H. 1995, ApJ, 445, 433

\bibitem[1996]{allard96} Allard, F., Hauschildt, P. H., Baraffe, I., 
         Chabrier, G. 1996, ApJ, 465, L123

\bibitem[1997]{allard97} Allard, F., Hauschildt, P. H., Alexander, D. R., 
         Starrfield, S. 1997, ARA\&A, 35, 137

\bibitem[1999]{basri99} Basri, G., et al. 1999, ApJ, in press

\bibitem[1999]{burgasser99} Burgasser, et al. 1999, ApJ, 522, L65

\bibitem[1977]{cymbal77} Cymbal, V. V. 1977, in Tables of the Franc-Condon 
         factors with account of vibraton-rotational interaction for 
         astrophysically important molecules. I. Titanium oxide,  
         deposited N 246--77

\bibitem[1997]{delfosse97} Delfosse, X., Tinney, C. G., Forveille, T., 
         et al. 1997, A\&A, 327, L25

\bibitem[1982]{gurvitz82} Gurvitz, L. V., Weitz, I. V., Medvedev, V. A. 1982, 
       Thermodynamic properties of individual substances. Moscow.
       Science

\bibitem[1979]{huber79} Huber, K. P., Herzberg G. 1979, in Constants of 
         Diatomic Molecules, Van Nostrand Reinhold, N. Y.

\bibitem[1991]{irwin91} Irwin, M., McMahon, R. G., Reid, N. 1991, MNRAS,
         252, 61

\bibitem[1998]{israelian98} Israelian, G., Garc\'\i a L\'opez, R., Rebolo, 
         R. 1998, ApJ, 507, 805

\bibitem[1996]{jones96} Jones, H. R. A., Longmore, A. J., Allard, F., 
         Hauschildt, P. H. 1996, MNRAS, 280, 77

\bibitem[1997]{jones97} Jones, H. R. A., \& Tsuji, T. 1997, ApJ, 480, L39

\bibitem[1999b]{kirk99b} Kirkpatrick, J. D., Allard, F., Bida, T., 
         Zuckerman, B., Becklin, E. E., Chabrier, G., Baraffe, I. 1999b,
         ApJ, 519, 834

\bibitem[1999a]{kirk99a} Kirkpatrick, J. D., Reid, I. N., Liebert, J., 
         et al. 1999a, ApJ, 519, 802

\bibitem[1993]{kurucz93} Kurucz, R. L. 1993, CD ROM \#9, Cambridge, MA: 
         Smithsonian Astrophysical Observatory

\bibitem[1998]{leggett98} Leggett, S. K., Allard, F., Hauschildt, P. H. 
         1998, ApJ, 509, 836

\bibitem[1989]{lunine89} Lunine, J. I., Hubbard, W. B., Burrows, A., 
         Wang, Y.-P., Garlow, K. 1989, ApJ, 338, 314

\bibitem[1993]{magazzu93} Magazz\`u, A., Mart\'\i n, E. L., Rebolo, R. 1993, 
         ApJ, 404, L17

\bibitem[1996]{marley96} Marley, M. S., Saumon, D., Guillot, T., Freedman, 
         R. S., Hubbard, W. B., Burrows, A., Lunine, J. I. 1996, Science, 
         272, 1919

\bibitem[1997a]{martin97a} Mart\'\i n, E. L., Basri, G., Delfosse, X., 
         Forveille, T. 1997a, A\&A, 327, L29

\bibitem[1999]{martin99} Mart\'\i n, E. L., Delfosse, X., Basri, G., 
         Goldman, B., Forveille, T., Zapatero Osorio, M. R. 1999,
         AJ, in press (November issue)

\bibitem[1997b]{martin97b} Mart\'\i n, E. L., Pavlenko, Ya. V., ,Rebolo, R. 
         1997b, A\&A, 326, 731

\bibitem[1996]{martin96} Mart\'\i n, E. L., Rebolo, R., Zapatero Osorio, 
         M. R. 1996, ApJ, 469, 706

\bibitem[1996]{matthews96} Matthews, K., Nakajima, T., Kulkarni, S. R., 
         Oppenheimer, B. R. 1996, AJ, 112, 1678

\bibitem[1995]{nakajima95} Nakajima, T., Oppenheimer, B. R., Kulkarni, S. 
         R., Golimowski, D. A., Matthews, K., Durrance, S. T. 1995, 
         Nature, 378, 463

\bibitem[1995]{oke95} Oke, J. B., et al. 1995, PASP, 107, 375

\bibitem[1995]{oppenheimer95} Oppenheimer, B. R., Kulkarni, S. R., 
         Matthews, K., Nakajima, T. 1995, Science, 270, 1478

\bibitem[1996]{osterbrock96} Osterbrock, D. E., Fulbright, J. P., Martel, 
         A. R., Keane, M. J., Trager, S. C., Basri, G. 1996, PASP, 108, 277

\bibitem[1997]{pav97} Pavlenko, Ya. V, 1997, Astron. Reports, 41, 537

\bibitem[1998]{pav98} Pavlenko, Ya. V. 1998, Astron. Reports, 42, 787

\bibitem[1999a]{pav99a} Pavlenko, Ya. V. 1999a, Astron. Reports, 43, 115

\bibitem[1999b]{pav99b} Pavlenko, Ya. V. 1999b, Astron. Reports, in press

\bibitem[1995]{pav95} Pavlenko, Ya. V., Rebolo, R.,  Mart\'\i n, E. L,
         Garc\'\i a L\'opez, G. A\&A, 303, 807.

\bibitem[2000]{pav+oso+reb99} Pavlenko, Ya. V., Zapatero Osorio, M. R., 
         Rebolo, R. 2000, in Low-Mass Stars and Brown Dwarfs in Stellar 
         Clusters and Associations, La Palma, CUP, in press

\bibitem[1995]{piskunov95} Piskunov, N. E., Kupka, F., Ryabchikova, T. A., 
         Weiss, V. V., Jeffery, C.S. 1995, A\&AS , 112, 525

\bibitem[1992]{rebolo92} Rebolo, R., Mart\'\i n, E. L., Magazz\`u, A. 1992,
         ApJ, 389, L83

\bibitem[1996]{rebolo96} Rebolo, R., Mart\'\i n, E. L., Basri, G., 
         Marcy, G. W., Zapatero Osorio, M. R. 1996, ApJ, 469, L53

\bibitem[1998]{rebolo98} Rebolo, R., Zapatero Osorio, M. R., Madruga, S., 
         B\'ejar, V. J. S., Arribas, S., Licandro, J. 1998, Science, 282,
         1309

\bibitem[1997]{ruiz97} Ruiz, M. T., Leggett, S. K., Allard, F. 1997, ApJ,
         491, L107

\bibitem[1998]{schwenke98} Schwenke, D. 1998, in Chemistry and Physics of 
         Molecules and Grains in Space. Faraday Discussions No. 109. The 
         Faraday Division of the Royal Society of Chemistry, London, 
         p. 321

\bibitem[1998]{schultz98} Schultz, A. B., Allard, F., Clampin, M., et al. 
         1998, ApJ, 492, L181

\bibitem [1969]{schumaker69} Schumaker, J.B. 1969. JQSRT, v. 9, p. 153

\bibitem[1999]{strauss99} Strauss, M. A., et al. 1999, ApJL, 522, L61

\bibitem[1998]{tinney98} Tinney, C. G. 1998, MNRAS, 296, L42

\bibitem[1993]{tinney93} Tinney, C. G., Mould, J. R., Reid, I. N. 1993, 
         AJ, 105, 1045

\bibitem[1999]{tokunaga99} Tokunaga, A. T., Kobayashi, N. 1999, AJ, 117, 1010

\bibitem[1973]{tsuji73} Tsuji, T. 1973, A\&A, 23, 411

\bibitem[2000]{tsuji00} Tsuji, T. 2000, in Low-Mass Stars and Brown Dwarfs 
         in Stellar Clusters and Associations, La Palma, CUP, in press

\bibitem[1996a]{tsuji96a} Tsuji, T., Ohnaka, K., Aoki, W. 1996a, A\&A, 305, 1

\bibitem[1999]{tsuji99} Tsuji, T., Ohnaka, K., Aoki, W. 1999, ApJ, 520, 
         L119

\bibitem[1996b]{tsuji96b} Tsuji, T., Ohnaka, K., Aoki, W., Nakajima, T. 
         1996b, A\&A, 308, 29

\bibitem[1998]{yakovina98} Yakovina, L. A., Pavlenko, Ya. V. 1998,  
         Kinemat. and Phys. of Celest. Bodies, 14, 195



\end{thebibliography}
\end{document}